# Analysis of Demand Driven Ad Hoc Routing Protocols on Performance and Mobility

Biju Issac, Khairuddhin Ab Hamid and C. E. Tan
University Malaysia Sarawak (UNIMAS), Malaysia.

*Abstract* – **Mobile ad hoc networking (MANET) is a growing technology that can support the operation of adaptive wireless networks. With the increased demand rate of wireless applications it is useful to have more adaptive and self-organizing technologies that adapt to changes within a network region. In this paper we initially present a brief listing of table driven ad hoc routing protocols and eventually analyze in detail the behavior of demand driven ad hoc routing protocols like – Ad-Hoc On-Demand Distance Vector Routing (AODV), Dynamic Source Routing (DSR), Temporally-Ordered Routing Algorithm (TORA) in terms of throughput, traffic dropped, routing traffic and mobility. The output graphs are eventually discussed, thus enabling us to understand the routing technology better.**

## I. INTRODUCTION

The mobile wireless network can be broadly classified into two modes, namely – infrastructure mode and ad hoc mode. In infrastructure mode, a central base station is used for the mobile nodes to communicate with as long as it is within the communication radius of base station. Handover happens when the mobile node moves into the communication radius of another base station, leaving the previous one. In ad hoc mode, there are no fixed routers and all nodes are capable of movement and can be connected in a dynamic and arbitrary way. Nodes of these network act as routers that discover and maintain routes to other nodes. Example applications of ad hoc networks can be emergency search and rescue operations and data acquisition operations in unwelcoming terrains. Ad hoc routing protocols are getting popular with the increase in mobile computing. Ad hoc networks include resource-starving devices, low bandwidth, high error rates and a topology that is continuously changing. Some of the design goals with ad hoc routing protocols are – minimal control overhead, minimal processing overhead, multi-hop routing capability, dynamic topology maintenance and loop prevention. The protocols should operate in a distributed manner. The nodes should operate either in proactive or reactive mode. This can also be termed as table based or demand-driven modes respectively. Proactive protocols are table-based and maintain routes for the entire network within each node. The nodes must be fully aware of the changing topology. The table based mode is quite an old way to get routing done in ad hoc networks. By keeping routing tables to store the location information of other nodes, this approach distributes the routing data to other nodes. For topologies that are overtly dynamic, the above approach can introduce a considerable overhead. Reactive protocols or demand driven protocols trade off this overhead with increased delay. A route to destination is established when it is needed based on an initial discovery between the source and destination [1]. In our paper, we analyze the different demand driven ad-hoc routing protocols' performance and other behavior. The paper is organized as follows. Section II gives a brief analysis of mobile ad hoc table-driven protocols, Section III gives a comparison of mobile ad hoc demand-driven protocols, Section IV shows the simulation result with demand-driven protocols, Section V gives the comparison summary on demand-driven protocols, Section VI is related work and Section VII is the conclusion.

## II. TABLE DRIVEN PROTOCOLS

In table driven approach, the freshness of the routing tables is ensured by broadcasting a hello packet with address information at frequent intervals of time. Each node thus updates the location information of other participating nodes. *Destination-Sequenced Distance Vector (DSDV), Clusterhead Gateway Switch Routing (CGSR)* and *Wireless Routing Protocol (WRP)* are some examples of table driven protocols. A brief description of different table driven routing protocols is given below: *Destination-Sequenced Distance Vector* is a different version of the Bellman-Ford algorithm to take care of time dependent topologies. Each node maintains a routing table with the next hop entry for each destination and metric for each link. Each entry is marked with a sequence number assigned by the destination node, which enables the mobile nodes to distinguish stale routes from new ones, thus avoiding the routing loops. The overhead with this protocol is high and this limits the number of nodes in the network. In order to reduce the packet overhead, two types of packets are used here – full dump and incremental packets. Full dump packets which contain all available routing information are sent infrequently. Incremental packets are used to relay only the changes since the last full dump packet and are sent frequently. *Clusterhead Gateway Switch Routing* differs from DSDV by the type of addressing and networking scheme employed. A clusterhead node or station is selected here using a cluster head selection algorithm to control a group of ad hoc nodes. A Least Cluster Change (LCC) clustering algorithm is executed each time when two cluster heads come into contact or when a node moves out of contact of all other cluster heads, instead of invoking the clusterhead reselection algorithm. It modifies DSDV approach by using a hierarchical clusterhead-to-gateway routing architecture to route traffic from source to destination. Gateway nodes are those nodes that are within the communication range of two or more cluster heads. A packet sent by a source station is routed to its clusterhead and then to the gateway to another clusterhead and so on until the cluster head of the destination station is reached. It is then directed to the destination station. Each node maintains a cluster member table and a routing table to get the details to reach the destination. The *Wireless Routing Protocol* is designed to maintain routing information among all nodes in the network. Four tables are maintained by all nodes, namely – Distance table, Routing table, Link-cost table and Message retransmission list (MRL) table. Through the update messages that are sent between neighboring nodes, the mobile nodes inform each other of the link changes. It contains a list of updates such as the destination, the distance to destination and the node before the destination. The neighbors then update their distance table entries and check for new possible paths, which are relayed back to the original nodes so that their tables can be updated. The nodes maintain connectivity by sending *hello* messages at definite intervals of time. The new nodes that send *hello* messages are updated into mobile node's routing table and the mobile node sends a copy of its routing information to them. A brief comparison is shown in table I [2].

TABLE I
TABLE-DRIVEN PROTOCOL COMPARISON

| Parameters | DSDV | CGSR | WRP |
|---|---|---|---|
| Routing philosophy | Flat | Hierarchical | Flat |
| No. of required tables | Two | Two | Four |
| Use Sequence Numbers | Yes | Yes | Yes |
| Use hello packets | Yes | No | Yes |
| Routing metric (path) | Shortest | Shortest | Shortest |
| Loop-free network | Yes | Yes | Yes (takes some time) |

## III. DEMAND DRIVEN PROTOCOLS

A different approach in comparison to the table driven routing is the source initiated demand driven protocols that creates the route only when needed by the source node. When a route is needed by a source node, it would initiate a route discovery process within the network. Once a route is discovered and established, it is maintained until it is needed. *Ad-Hoc On-Demand Distance Vector Routing (AODV)* is a variation of DSDV algorithm. As a reactive protocol, the finding of route is based on a route discovery cycle that includes a network search that is broadcasted and a unicast reply containing the discovered paths. When a node wants to establish a communication link, it does a path-discovery process to locate the other node. If a path to destination exists in the source node's routing table, it will use that existing route to send data packets. Otherwise to start the route discovery, the source node broadcasts a *route request*, RREQ packet with it's and destination's IP addresses, Broadcast ID (or RREQ ID) and the sequence numbers of source and destination nodes. The RREQ ID is a per-node counter that is incremented every time the node initiates a new RREQ. This makes the RREQ ID together with the source IP address quite unique and can be used to identify a particular RREQ. When a node receives this RREQ, it creates a reverse path to the node and the hop count in the RREQ is incremented by one. If the current node does not have an unexpired route to destination the RREQ is broadcasted to its neighbors with incremented hop count value, thus creating a flooding scenario. Once the receiving node's sequence number is greater than or equal to the destination station's sequence number as indicated in the RREQ, it generates a *route reply*, RREP (reply message) indicating the route. The node then unicasts the message to its next hop towards source node. Thus the receiving nodes set a backward pointer to the source and generate a RREP unicast packet. As RREP is routed back to source, intermediate nodes set up forward pointers in their routing tables. The RREP packets contain the source and destination IP addresses, destination's sequence number etc. If the destination itself is creating the RREP, the hop count is made as zero. The reverse route created as RREQ is forwarded and is used to channel the RREP back to the source. AODV nodes store a route table which stores the next-hop routing information with a designated lifetime for destination nodes. If the routes are used, the lifetime is extended. The route expires after the lifetime, if not used. AODV favors the least congested route and uses *hello* messages to maintain node connectivity.

*Dynamic Source Routing (DSR)* also uses source based routing rather than table based and it is source initiated rather than hop-by-hop. A node wishing to communicate issues a Route Request to all its neighbors and each neighbor in turn rebroadcasts this Request adding its own address in the header of the packet. When a Request is received by the destination node or by an intermediate node in the path to destination, a Route Reply is sent back to the sender along with addresses accumulated in the Request header. Thus the entire route is maintained in the packet header. DSR is not that scalable to large networks.

*Temporally-Ordered Routing Algorithm (TORA)* is a distributed and adaptive protocol that can operate in a dynamic network. For a particular destination, it uses the 'height' parameter to determine the direction of a link between any two nodes. Thus multiple routes are often present for a given destination. For a node to start communication, it broadcasts a Query to all its neighbors, which is rebroadcast through the network until it reaches the destination. This node would reply with an Update that includes its height with respect to the destination, which is sent back to the sender. Each node that receives the Update sets its own height to one greater than that of the neighbor that sent it, which results in a series of directed links from sender to destination in the order of height values. Internet MANET Encapsulation Protocol (IMEP) is normally run at the layer below TORA so that some functionalities from lower layers can be used by TORA [3]-[6].

## IV. SIMULATION OF DEMAND-DRIVEN PROTOCOLS

We did a network simulation of existing demand driven protocols like DSR, TORA and AODV [3], [5]-[6] to study the performance and to analyze its behavior, under different number of nodes and other changes in parameters, using existing models. For DSR, the scenarios created could compare for example, the throughput, total routing traffic, dropped traffic etc. For TORA also, the scenarios compare the throughput, total routing traffic and dropped traffic. For AODV, its performance is first analyzed with default parameters. It is then compared to a scenario where traffic is decreased by tuning some parameters. Our main focus would be on the above performance analysis. The mobility aspects under two protocols – DSR and AODV would also be investigated briefly. In DSR scenario, all nodes in the network are configured to have mobility by configuring trajectories. Every node runs DSR and multiple FTP or Telnet sessions. In AODV also, all nodes in the network are configured to have mobility by configuring trajectories. Output comparison graphs would be discussed to analyze the protocol behavior, under different circumstances.

*A. Performance Comparison*

As shown in figure 1, firstly a wireless server is configured with DSR and 40 mobile nodes are configured as work stations (that can generate application traffic) running DSR. They are placed as 20 nodes on one side and remaining 20 nodes on the other side. Later the same exercise is done with 80 nodes, by placing 40 nodes each on either side of the server. The simulation is done for 30 minutes (1800 sec).

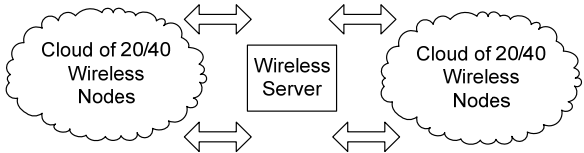

Fig.1 The general mobile ad hoc network topology used for simulation.

The wireless nodes can communicate to each other within a cloud or to the server. They can also communicate with the wireless nodes in the other cloud. Figure 2 shows the generated throughput of 40 and 80 node DSR networks that run some applications like Telnet, Email or FTP sessions. The mobile nodes are configured to eliminate all receivers that are over 1500 meters away.

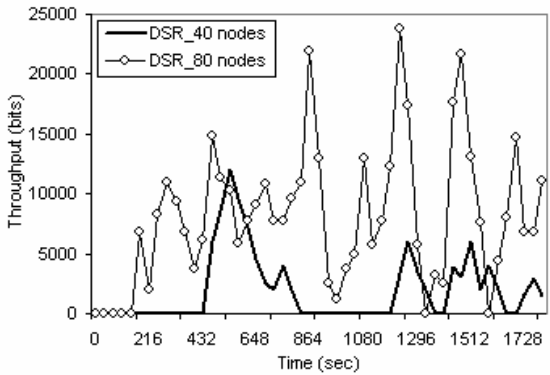

Fig.2 The throughput of ad-hoc network when using DSR with 40 nodes and 80 nodes.

The graph shows an expected result that the throughput of 80 node DSR network is greater than the 40 node DSR network, reaching a peak of around 24,000 bits (during 1200 sec) compared to 12,000 bits (during 500 sec) for 40 nodes. Lack of scalability of DSR, in terms of functioning well in bigger networks is evident from the performance graphs as it is meant for low diameter ad hoc mobile networks.

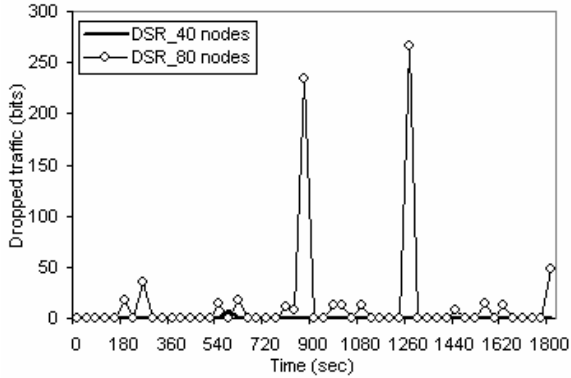

Fig.3 The amount of traffic dropped for ad-hoc network when using DSR with 40 nodes and 80 nodes. Note the high drop for 80 nodes.

As the traffic gets higher for 80 nodes, the traffic dropped also hits higher compared to 40 nodes, reaching a peak of around 270 bits (during 1200 sec) as shown in figure 3. For 40 nodes DSR network, the dropped traffic is almost negligible.

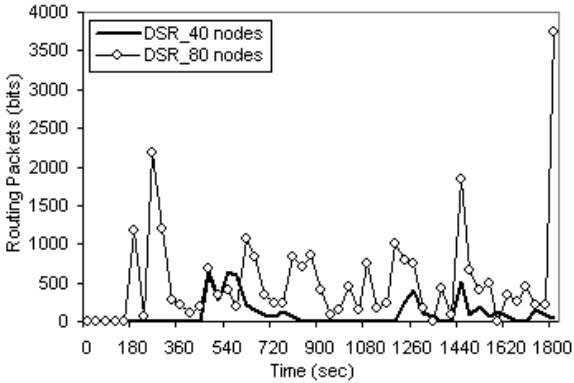

Fig.4 The routing packets sent (received is similar) for ad-hoc network when using DSR with 40 nodes and 80 nodes.

The routing traffic sent and received, as expected would be more for 80 nodes DSR network, reaching a peak of around 3800 bits during 30 minutes. This happens because the route request is flooded to all its neighbors (which in turn broadcasts it) when a node wants to communicate. So higher the number of nodes, higher the traffic generated.

Secondly, a wireless server is configured with TORA and 40 mobile nodes are configured as work stations running TORA by placing 20 nodes on one side and remaining 20 nodes on the other side. Later the same exercise is done with 80 nodes, by placing 40 nodes each on either side of the server. The simulation is done for 30 minutes (1800 sec).

Figure 5 shows the generated throughput of 40 and 80 node TORA networks that run some applications like Telnet, Email or FTP sessions. The nodes are configured to eliminate all receivers that are over 1500 meters away. In contrast to DSR, TORA works better with higher number of nodes as it is adaptive and works well dynamically with 80 nodes and the throughput graph for 40 nodes and 80 nodes TORA networks looks comparable.

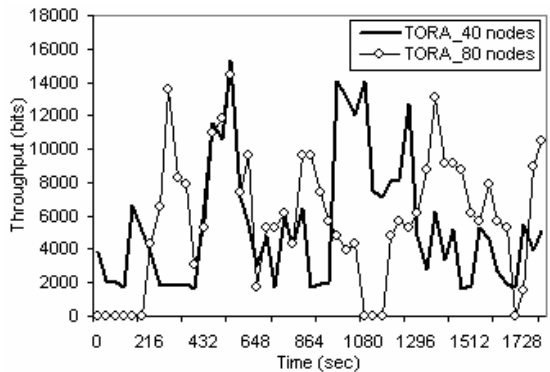

Fig.5 The throughput of ad-hoc network when using TORA with 40 nodes and 80 nodes.

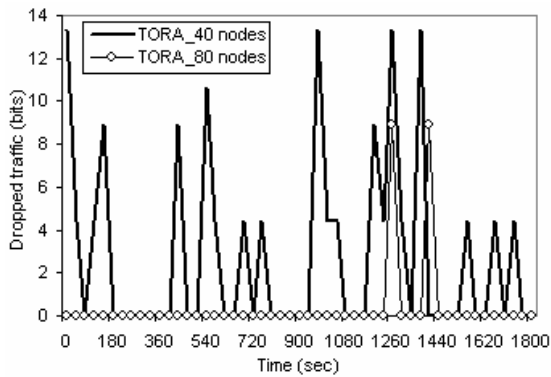

Fig.6 The amount of traffic dropped for ad-hoc network when using TORA with 40 nodes and 80 nodes. Note the high drop for 40 nodes.

As in figure 6, the traffic drop is negligible with 80 nodes TORA network and 40 nodes network also gives marginally low dropped traffic with a maximum of 13 bits. Generally the drop in traffic looks better than DSR in comparison with the scenarios of 40 and 80 nodes discussed before, where 80 nodes DSR shows few peaks of traffic drops. Figure 7 shows the routing packets sent and received using TORA.

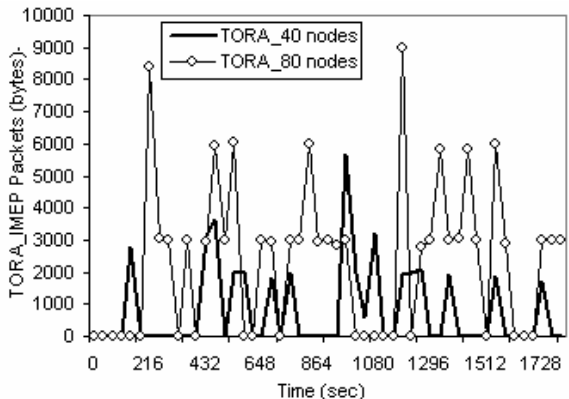

Fig.7 The routing packets sent (received is similar) for ad-hoc network when using TORA with 40 nodes and 80 nodes.

Thirdly, a wireless server is configured with AODV and 40 mobile nodes are configured as work stations running AODV by placing 20 nodes on one side and remaining 20 nodes on the other side. This is done in the default mode. Later the same exercise is done with 40 nodes, with reduced or light traffic. The simulation is done for 30 minutes (1800 sec). Figure 8 shows the generated throughput of 40 node AODV network that runs some applications like Telnet, Email or FTP sessions. It is configured to eliminate all receivers that are over 1500 meters away. Parameters set for AODV (to reduce routing traffic) are as follows: Route Discovery Parameter: Gratuitous Reply – Enabled, Active Route Timeout: 30 sec and WLAN data rate: 1Mbps.

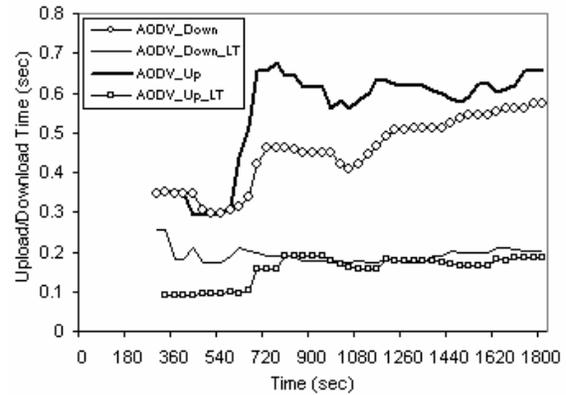

Fig.8 The FTP upload and download time for ad-hoc network when using AODV with 40 nodes. The default setting and Less Traffic (LT) setting is used.

Figure 9 shows routing traffic sent and received when using AODV in default and light traffic modes. Obviously, the routing traffic curves are low in the case of light traffic scenario where routing throughput is low.

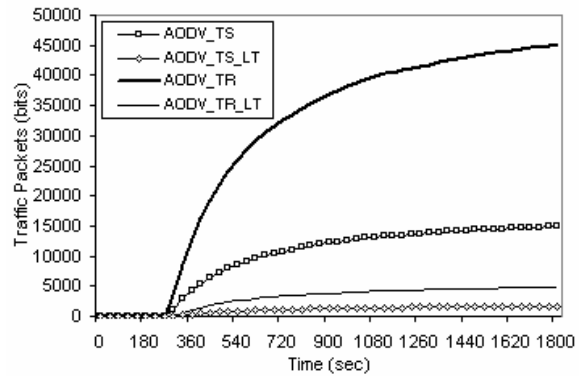

Fig.9 The routing traffic sent (TS) and traffic received (TR) for ad-hoc network when using AODV with 40 nodes. The default setting and Less Traffic (LT) setting is used.

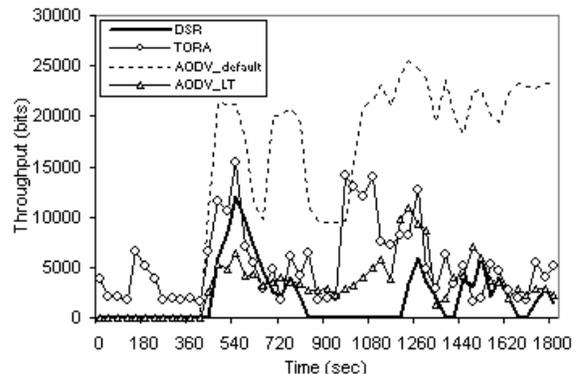

Fig.10 The Throughput for ad-hoc network when using DSR, TORA and AODV (default and low traffic (LT)) with 40 nodes.

Figure 10 shows the throughput comparison of all 3 protocols –DSR, TORA and AODV. AODV with gratuitous reply enabled (light traffic/LT option) and DSR shows comparable throughput. Default AODV shows a greater throughput and TORA is in between.

*B. Mobility Comparison*

Figure 11 shows the topology used for the study. Here 20 mobile nodes (with raw traffic generator over IP or application traffic) are made mobile by configuring trajectories that make them move left or right in the upward or downward direction.

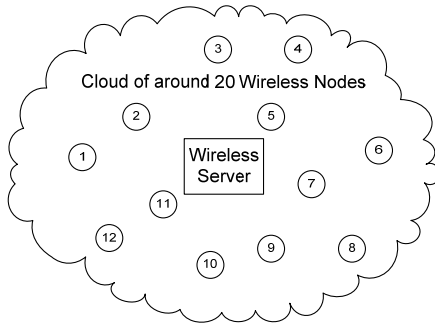

Fig.11 The network topology used in mobility analysis. A wireless server was surrounded by a cloud of around 20 mobile nodes which had trajectories defined. Only 12 nodes are shown in the diagram.

Figure 12 shows the traffic in a 20 node DSR network. All nodes in the network by configuring trajectories are set to have mobility and every node runs some applications like HTTP, Telnet or FTP sessions. Nodes are configured to eliminate all receivers that are over 500 meters away. Later it is changed to 1000 meters elimination. Since the network is mobile, the refresh interval is set to 10 seconds. The total traffic sent peaks to 40,000 – 45,000 bits as shown in the graph. WLAN data rate was 1 Mbps.

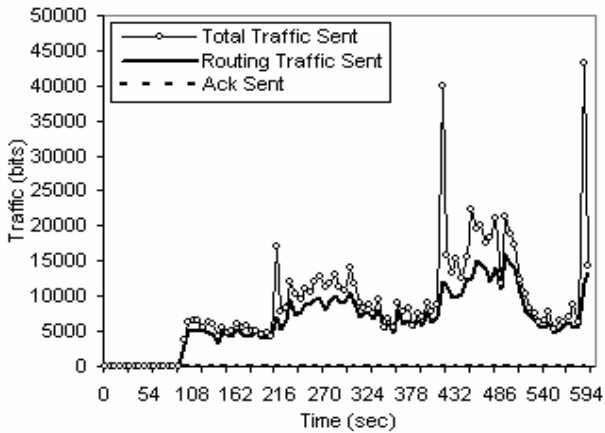

Fig.12 The traffic pattern for ad-hoc network when using DSR with mobility on multiple nodes. The sending and receiving patterns are similar.

Figure 13 shows the traffic drop when the receivers are eliminated over 500 meters away and 1000 meters away. Clearly, 500 meter elimination shows greater drop in traffic. Also the traffic dropped with DSR network averages around 130 packets for 500 meter elimination as shown in figure 12 for the topology and parameters chosen. Figure 14 shows the traffic in a 20 node AODV network with mobility by configuring trajectories. It is configured to eliminate all receivers that are over 1500 meters away. Since the network is mobile, the refresh interval is set to 5 seconds.

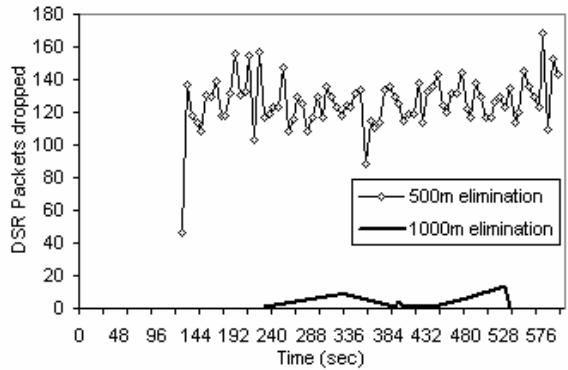

Fig.13 The DSR packets dropped for 2 cases (500m elimination & 1000m elimination) when using DSR with mobility on multiple nodes.

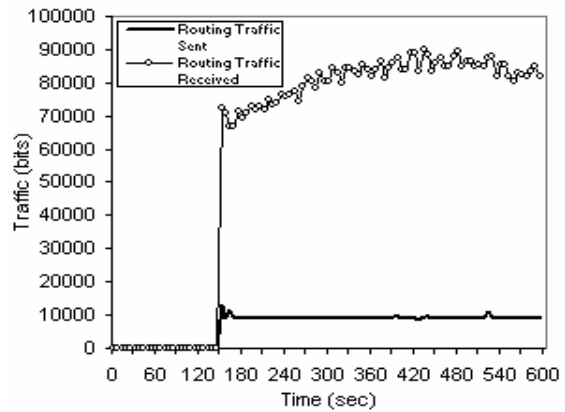

Fig.14 The traffic pattern for ad-hoc network when using AODV with mobility on multiple nodes.

Figure 15 shows that the traffic dropped is comparatively low in AODV with 500 meter and 1000 meter receiver elimination.

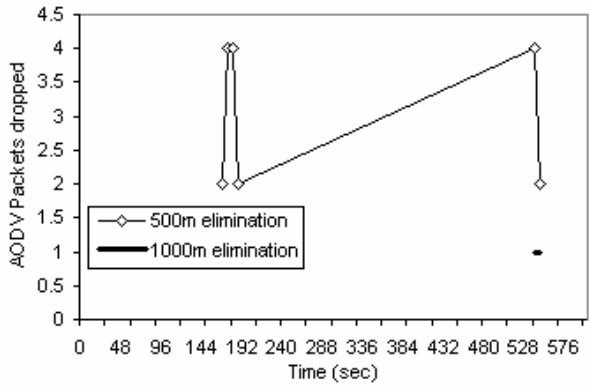

Fig.15 The AODV packets dropped for 2 cases (500m elimination & 1000m elimination) when using AODV with mobility on multiple nodes.

V. SUMMARY ON DEMAND-DRIVEN PROTOCOL FEATURES AND PERFORMANCE

A comparison summary on the three simulated protocols can be given as follows. The table 2 shows [2] the details in a nutshell. DSR is meant for wireless ad hoc networks where the mobile nodes move with moderate speed in comparison to packet transmission latency [7]. As DSR packets need to contain full routing information, the memory overhead is more compared to AODV. But it does not make use of periodic router advertisement which saves some bandwidth and power consumption. A direct result of this is that router advertisement overhead would be nil and topology changes are eliminated. As DSR allows multiple routes to the destination, in the event of a link failure, the other valid routes can be checked and this can prevent route reconstruction, hastening the route recovery. If multiple routes are not available, route discovery should be done to find a route. As DSR assumes small network diameter, it is not scalable to large networks. The requirement to place the route in both route replies and data packets increases the control overhead much more than in AODV.

TABLE II
DEMAND-DRIVEN PROTOCOL COMPARISON

| Parameters | AODV | DSR | TORA |
|---|---|---|---|
| Routing philosophy | Flat | Flat | Flat |
| Routes stored in | Route Table | Route Cache | Route Table |
| Multicast option | Yes | No | No |
| Multiple routes | No | Yes | Yes |
| Routing metric (path) | Most Fresh and Shortest | Shortest | Shortest |
| Loop-free network | Yes | Yes | Yes |
| Route re-configuration approach | Delete route – and notify source | Delete route – and notify source | Link reversal – and route repair |

AODV overhead is much lesser in comparison to DSR as the route replies need to carry only the destination IP address and sequence number. This also reduces the memory overhead in comparison to DSR. AODV also uses a route discovery mechanism that is similar to DSR. A big advantage of AODV is its support for multicast communication. But on the negative side it needs symmetric links between nodes and cannot make use of asymmetric links. TORA is a link-reversal algorithm that is suitable for densely populated network with large number of nodes. It creates directed acyclic graphs (DAG) to help route creation and supports multiple routes for single source-destination pair. Multiple routes reduce route reconstruction overheads. This protocol supports multicast in conjunction with Lightweight Adaptive Multicast Algorithm (LAM) by working as an underlying protocol [8]. TORA needs synchronized clocks for its operation and if the external time source fails, the protocol would fail. Even route rebuilding can incur lengthy delays because of synchronization related oscillations [2], [4]-[6].

VI. RELATED WORK

Broch, et. al. evaluated four ad hoc routing protocols including AODV and DSR. They used only 50 node models and traffic loads were kept low – around 4 packets/sec, 10–30 sources and 64 byte packets. Packet delivery fraction, number of routing packets and distribution of path lengths were used as performance metrics [9]. Johansson, et. al. extended the above work by using new mobility models. To characterize these models, a new mobility metric is introduced that measures mobility in terms of relative speeds of the nodes rather than absolute speeds and pause times. Again, only 50 nodes were used. A limited amount of load test was performed, but the number of sources was always small [10]. Another related work is by Das et. al. [11].

VII. CONCLUSION

We were able to do the performance and mobility analysis on the three demand driven ad-hoc routing protocols, namely – Ad Hoc On-Demand Distance Vector Routing (AODV), Dynamic Source Routing (DSR) and Temporally-Ordered Routing Algorithm (TORA). The performance and mobility graphs were discussed and found that the different protocol's performance varies depending on different parameters like lesser or higher number of nodes, light or heavy traffic, speed of mobile nodes etc. In comparison to the table driven protocols, the demand driven protocols will have to wait until a new route is discovered.

Biju Issac is a part-time PhD scholar in University Malaysia Sarawak (UNIMAS), Malaysia and lecturer in Swinburne University of Technology (Sarawak Campus), Kuching, Malaysia. His email address is bissac@swinburne.edu.my. Prof. Khairuddhin Ab Hamid (Deputy Vice Chancellor) and Dr.C. E. Tan who is in the Faculty of Computer Science and Information Technology are the co-authors who are also working in UNIMAS, Malaysia.